\documentclass[aps,prb,amsmath,amssymb,twocolumn,showpacs,floatfix,reprint,superscriptaddress,nobibnotes]{revtex4-1}

\usepackage[utf8]{inputenc} 
\usepackage{graphicx}
\usepackage{dcolumn}
\usepackage{bm}
\usepackage{units}
\usepackage{amsmath}
\usepackage{times}
\usepackage{miller}
\usepackage[caption=false]{subfig}
\usepackage{xcolor}
\graphicspath{{./}}

\newcommand{\dg}{$^\circ$}
\newcommand{\sub}[1]{\textsubscript{#1}}

\begin{document}

\title{The Basic and the Charge Density Wave Modulated Structures of NbS\sub{3}-II}

\author{E. Zupanič}
\email{erik.zupanic@ijs.si}
\author{H. J. P. van Midden}
\author{M. van Midden}
\author{S. Šturm}
\affiliation{Jožef Stefan Institute, Jamova 39, SI-1000 Ljubljana, Slovenia}

\author{E. Tchernychova}
\affiliation{Department of Materials Chemistry, National Institute of Chemistry, Hajdrihova 19, SI-1000 Ljubljana, Slovenia}

\author{V. Ya. Pokrovskii}
\email{pok@cplire.ru}
\author{S. G. Zybtsev}
\author{V. F. Nasretdinova}
\author{S. V. Zaitsev-Zotov}
\affiliation{Kotel’nikov Institute of Radioengineering and Electronics of Russian Academy of Sciences (RAS), Mokhovaya 11-7, 125009 Moscow, Russia}

\author{W. T. Chen}
\affiliation{Center for Condensed Matter Sciences, National Taiwan University, Taipei, Taiwan, 106}

\author{Woei Wu Pai}
\email{wpai@ntu.edu.tw}
\affiliation{Center for Condensed Matter Sciences, National Taiwan University, Taipei, Taiwan, 106}
\affiliation{Department of Physics, National Taiwan University, Taipei, Taiwan 106}

\author{J. C. Bennett}
\affiliation{Department of Physics, Acadia University, Wolfville, Nova Scotia, Canada B4P 2R6}

\author{A. Prodan}
\affiliation{Jožef Stefan Institute, Jamova 39, SI-1000 Ljubljana, Slovenia}

\begin{abstract}
The basic and the charge density wave (CDW) structures of the monoclinic NbS\sub{3}-II polymorph were studied by synchrotron x-ray diffraction, ab-initio calculation, simulation of electron diffraction patterns and by atomic-resolution transmission electron and low-temperature scanning tunneling microscopies. It is shown that the basic structure belongs to the space group $P2_1/m$ and is described with a unit cell, formed of four pairs of symmetry-related trigonal prismatic (TP) columns ($a\sub{0}$ = 0.96509(8) nm, $b\sub{0}$ = 0.33459(2) nm, $c\sub{0}$ = 1.9850(1) nm, $\beta_0$ = 110.695(4)\dg), with all Nb and S atoms in $2e$ special positions. The two CDWs, with $\vec{q_1}$ = (0, 0.298,0) and $\vec{q_2}$ = (0, 0.352, 0), form their own modulation unit cells ($a\sub{m}$ = $2a\sub{0}$, $b\sub{m}$ = $b\sub{0}/q\sub{jb}$, $c\sub{m}$ = $c\sub{0}$, $\beta_m$ = $\beta_0$) and are ordered pairwise along adjacent isosceles TP columns. The symmetries of both $\vec{q\sub{j}}$ modes belong to the space group $Cm$ and are related according to the $2a$ special positions. If considered as long-period commensurate, the entire modulated structure with both CDWs included is best described with an enlarged unit cell ($a$ = $2a\sub{0}$, $b$ = $37b\sub{0}$, $c$ = $c\sub{0}$, $\beta$ = $\beta_0$), with all Nb and S atoms in $1a$ positions of the space group $P1$. 
 
\end{abstract}

\maketitle

\textit{Introduction.}--A large number of quasi one-dimensional (1D) transition-metal trichalcogenides (MX\sub{3}), which all possess related structures \cite{1,2}, are now known. They are composed of bi-caped trigonal prismatic (TP) columns, formed of chalcogen atoms and centered by metallic chains. The metal atoms are additionally linked to the chalcogen atoms of the neighboring columns, becoming thus eight-coordinated. The 1D columns are alternately shifted in phase by half the unit cell height and are strongly bonded into layers, separated by relatively wide Van der Waals (VdW) gaps. As a result the structures show a strong 1D as well as a two-dimensional (2D) character with very anisotropic structural and transport properties.  

Depending on the growth conditions, NbS\sub{3} single crystals present a variety of polymorphs\cite{new1}. However, only two of these (NbS\sub{3}-I and NbS\sub{3}-II) have been studied in detail \cite{2}. The structure of NbS\sub{3}-I was solved long ago \cite{2} and was characterized by a pronounced dimerization along its triclinic $\vec{b_0}$-axis. In contrast to NbS\sub{3}-I single crystals, whose dimensions are usually sufficiently large to enable conventional single-crystal x-ray structural analysis, single crystals of NbS\sub{3}-II are typically much smaller and have a less pronounced 2D character. The hair-like morphology, small size and tendency to bend of NbS\sub{3}-II crystals, coupled with a slight nonstoichiometry of about $3\%$ S deficiency, as measured by electron probe micro-analysis (EPMA)\cite{15}, has prevented the basic and the CDW-modulated structures of NbS\sub{3}-II from being accurately determined by means of single crystal x-ray analysis. It was however clearly shown in electron diffraction experiments that in the case of NbS\sub{3}-II weak pairs of satellites, whose incommensurate (IC) components appeared close to $\approx$ $\frac{1}{3}{b_0^*}$ in the corresponding diffraction patterns (DP)\cite{4,5}, replace the characteristic dimerization in NbS\sub{3}-I. The space groups of both the basic structure and the modulation were determined from electron DPs; the super-space group \cite{6,7} was determined as $A^{P2_1/m}_{~~1~~~~\bar{1}} \hspace{1mm} (no. 11b.6.1)$ and the corresponding dualistic notation \cite{8} as $^{mP2_1/m}_{~cB2~/m}$ (both with interchanged axes with regard to the present description, i.e. with $\vec{c_0}$ being the direction of the IC modulation) \cite{9}. Since many interesting and unique physical properties of NbS\sub{3}-II have been reported and summarized\cite{15}, it is imperative to determine the basic structure of NbS\sub{3}-II and rationalize the plausible CDW modulation structure as well. 

The unique structure-dependent physical properties of NbS\sub{3}-II, including the observation of a third independent CDW-driven phase transition \cite{15}, has rendered the identification of the heretofore unknown basic structure a problem of significant importance. As for a few other MX\sub{3} compounds, e.g. NbSe\sub{3} and the monoclinic polymorph of TaS\sub{3} (m-TaS\sub{3}) \cite{11b,23}, pairs of IC satellites in the electron DPs of NbS\sub{3}-II \cite{5} correspond to two CDWs with satellites at position (0.5, 0.298, 0) and (0.5, 0.352, 0) \cite{2,11,12}. The non-zero x-parameters (0.5) of these positions are in fact a result of the extinction rules, imposed by the modulation space group, while the actual modulation of both CDWs is directed along the columns with $\vec{q_1}$ = (0, 0.298,0) and $\vec{q_2}$ = (0, 0.352, 0). The two CDWs show different onset temperatures ($\vec{q_1}$ at T\sub{1} $\approx$ 360 K and $\vec{q_2}$ at T\sub{2} $\approx$ 600 K) and were shown to slide under an external electric field \cite{13}. A further interesting similarity between NbS\sub{3}-II and NbSe\sub{3} is the fact, that the two $\vec{q}$-vectors are in both cases related, with their IC components adding into a commensurate (COM) value within experimental error. If the structure is considered as long-period (LP) COM, the two IC CDWs in NbS\sub{3}-II fit closely to 5 and 6 modulation periods per 17${b_0}$ \cite{4} and could add closely to a COM value with $q_{1b} + q_{2b} \approx 2/3$ \cite{5}. However, the situation in NbS\sub{3}-II is further complicated in comparison to NbSe\sub{3}. In addition to the two aforementioned CDWs, which both show clear evidence of the structural modulation in electron DPs, a third CDW is detected in transport measurements, with an onset temperature of about 150 K \cite{13}. In contrast to the two previously identified CDWs, this third CDW shows a very low and vastly varied concentration of carriers (between 3 and 1000 times lower than that of $\vec{q_2}$) and practically no evidence of its existence in the corresponding structural studies \footnote{Except for an occasional and very weak detection in nuclear magnetic resonance (NMR) and x-ray absorption near edge structure (XANES) experiments\cite{15}}. Another complication in case of NbS\sub{3}-II is its appearance in two sub-phases, high-ohmic and low-ohmic samples \cite{13} with different electrical conductivities and possibly slight difference in stoichiometries. Only the low-ohmic sub phase shows the 150 K transition.

The present study was undertaken in order to acquire a better understanding of the basic and the modulation structures of NbS\sub{3}-II. Synchrotron x-ray powder diffraction (XRD), high-resolution (HR) scanning transmission electron (STEM) and scanning tunneling microscopy (STM) studies were performed in conjunction with ab-initio calculations of the basic structure, in relation to the known structures of NbS\sub{3}-I and NbSe\sub{3} and a comparison of the experimental and the calculated electron DPs. 

The employment of a broad range of complementary techniques has finally allowed a refinement of the NbS\sub{3}-II basic structure. In addition, the structure of both IC CDW modulations are solved and related to the basic structure. The details of the complete modulated structure, considered as LP COM, are also provided.\\

\textit{Experimental.}--The NbS\sub{3}-II whiskers were grown in a three-zone furnace with separate  temperature control for each zone with typical set temperatures of 750, 700, and 650\dg C \cite{13}.

HR STEM images were acquired with an aberration-corrected probe in a TEM (JEOL JEM-ARM200F), using a cold field emission source. The probe size was 0.1 nm, with a current of 20 pA and a convergence semi-angle of 24 mrad. The collection semi-angle for the high-angle annular dark-field (HAADF) detector was set between 45 and 180 mrad. Wiener filter was applied on high-resolution experimental images to improve the visibility of the crystalline matrix in the presence of partially amorphous surface layers \cite{temfilters}. The HAADF-STEM images were calculated using the QSTEM code \cite{18} by applying the same electron-optical parameters as used during the experimental image acquisition. Finally, the Poisson noise was added onto the calculated images to mimic the background noise of the HAADF detector. 

STM was conducted on an Omicron low-temperature (LT) STM operated between 78 and 160 K, with a base pressure of $2.5\cdot10^{-11}$ Torr. Flashed tungsten tips were used. The NbS\sub{3}-II whisker sampled were freshly cleaved in-situ in ultrahigh vacuum at room temperature.

XRD experiments were conducted for phase identification and for structural analysis. The XRD patterns were recorded with a Rayonix MX225HE CCD detector and a beam energy of 18 keV at the beamline BL12B2, Spring-8 (Japan), and with a MYTHEN detector and a beam energy of 15 keV at the beamline 09A, Taiwan Photon Source (TPS), National Synchrotron Radiation Research Center (Taiwan). The fine NbS\sub{3}-II whiskers were carefully and repeatedly cut by razor blades and packed into 0.3 mm borosilicate capillaries. The capillary was kept spinning during data collection for powder averaging. The diffraction patterns were analyzed with the Le Bail and Rietveld methods using the Bruker DIFFRAC.TOPAS \cite{topas} program. Since the Spring-8 and TPS data gave similar results, only the Spring-8 data are presented.

Density functional theory (DFT) calculations were performed using the Quantum-ESPRESSO code \cite{20}. In this approach the electron-ion interactions are described by the Garrity-Bennett-Rabe-Vanderbilt (GBRV) high throughput ultrasoft pseudopotentials \cite{21}, with the generalized gradient approximation of the exchange and correlation functional of Perdew, Burke, and Ernzerhof (PBE) \cite{22}. For the plane-wave basis set expansion, the kinetic cutoff energy was at 40 Ry. Based on the size of the simulation cell, either a (4,4,2) or a (2,4,2) Monckhorst-Pack k-point sampling grid was used. The atomic structure was relaxed until the changes in total energy were below $1.0 \cdot 10^{-6}$ Ry and the forces below $1.0 \cdot 10^{-5}$ atomic units.\\

\begin{figure}[ht!]
\centering
\captionsetup[subfloat]{labelformat=empty}
\subfloat[]{\includegraphics[width=\linewidth]{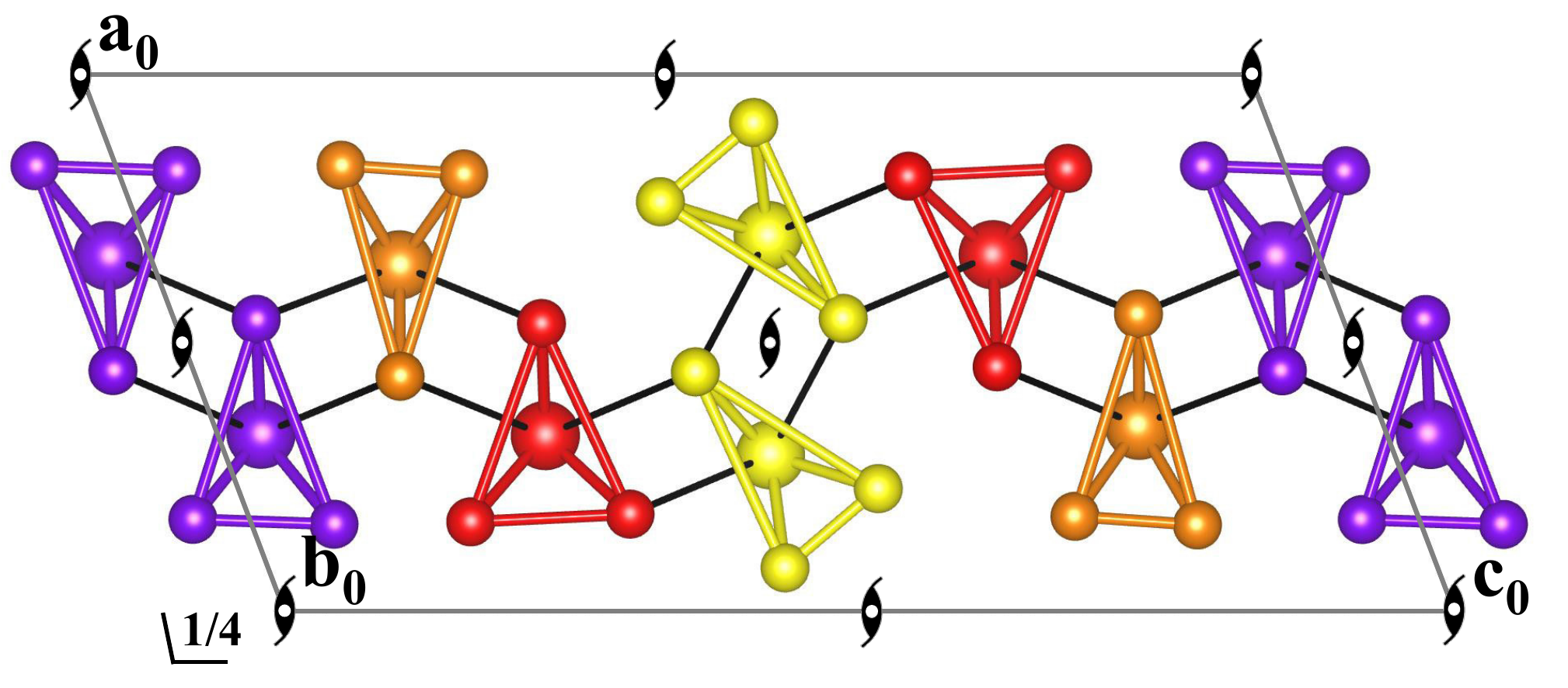}}
\caption{\label{models}The basic structure of NbS\sub{3}-II ($a\sub{0}$ = 0.96509(8) nm, $b\sub{0}$ = 0.33459(2) nm, $c\sub{0}$ = 1.9850(1) nm, $\beta_0$ = 110.695(4)\dg). Large balls represent Nb and the small ones S atoms. There are four symmetry-related pairs of trigonal prismatic columns in the unit cell, three isosceles (Y-yellow, O-orange and P-purple) and one almost equilateral (R-red). The two inter-column Nb-S bonds of the eight bonds forming the bi-capped trigonal prisms are shown black. The symmetry elements of the space group $P2_1/m$ are added.}
\end{figure}

\textit{Results.}--A main result is the successful refinement of the NbS\sub{3}-II basic structure, as shown in Figure~\ref{models}. Its unit cell is composed of four symmetry-related pairs of trigonal prismatic columns, three isosceles and one almost equilateral, which clearly resembles the structure of NbSe\sub{3}. The corresponding structural data are collected in Table~\ref{table1}.

In addition, the structures of both IC CDW modulations are solved and related to the basic structure. The structural details of both CDWs are shown in Figure~\ref{fig_modulation} and summarized in Table~\ref{table2}, while the details of the complete modulated structure, considered as LP COM, are given in Table~\ref{table3}. \\

\textit{Le Bail x-ray powder diffraction analysis.}--In comparison with XRD characterization using the Bragg-Brentano geometry, a significant improvement of diffraction profiles was achieved with synchrotron XRD. The Le Bail method was first applied to compare the experimental powder x-ray diagrams with several suggested unit-cell parameters \cite{new2,4,kikkawa,vansmaalen,10}. Best fit was obtained with the parameters of Van Smaalen\cite{vansmaalen}, $a\sub{0}$ = 0.9650 nm, $b\sub{0}$ = 0.3345 nm, $c\sub{0}$ = 1.8749 nm, $\beta_0$ = 98.07\dg, space group $P2_1/m$. However, we decided to use a transformed cell with $a\sub{0}$ = 0.9654 nm, $b\sub{0}$ = 0.3346 nm, $c\sub{0}$ = 1.9855 nm and $\beta_0$ = 110.71\dg in this work. These two unit-cells have identical cell volumes and the XRD patterns are equally well fitted with either of them. The transformed unit cell was chosen not only because the modified structure parameters clearly resemble that of NbSe\sub{3} but also because the two axes ($\vec{b\sub{0}}$ and $\vec{c\sub{0}}$) are chosen to be parallel to the VdW gaps. This offers a more straightforward characterization of the HR STEM and LT STM images as well as the corresponding electron DPs. It also resembles the known structure of NbSe\sub{3}\cite{11}.\\
\\

\begin{figure}[ht!]
\centering
\subfloat[fig1a][]{\includegraphics[width=\linewidth]{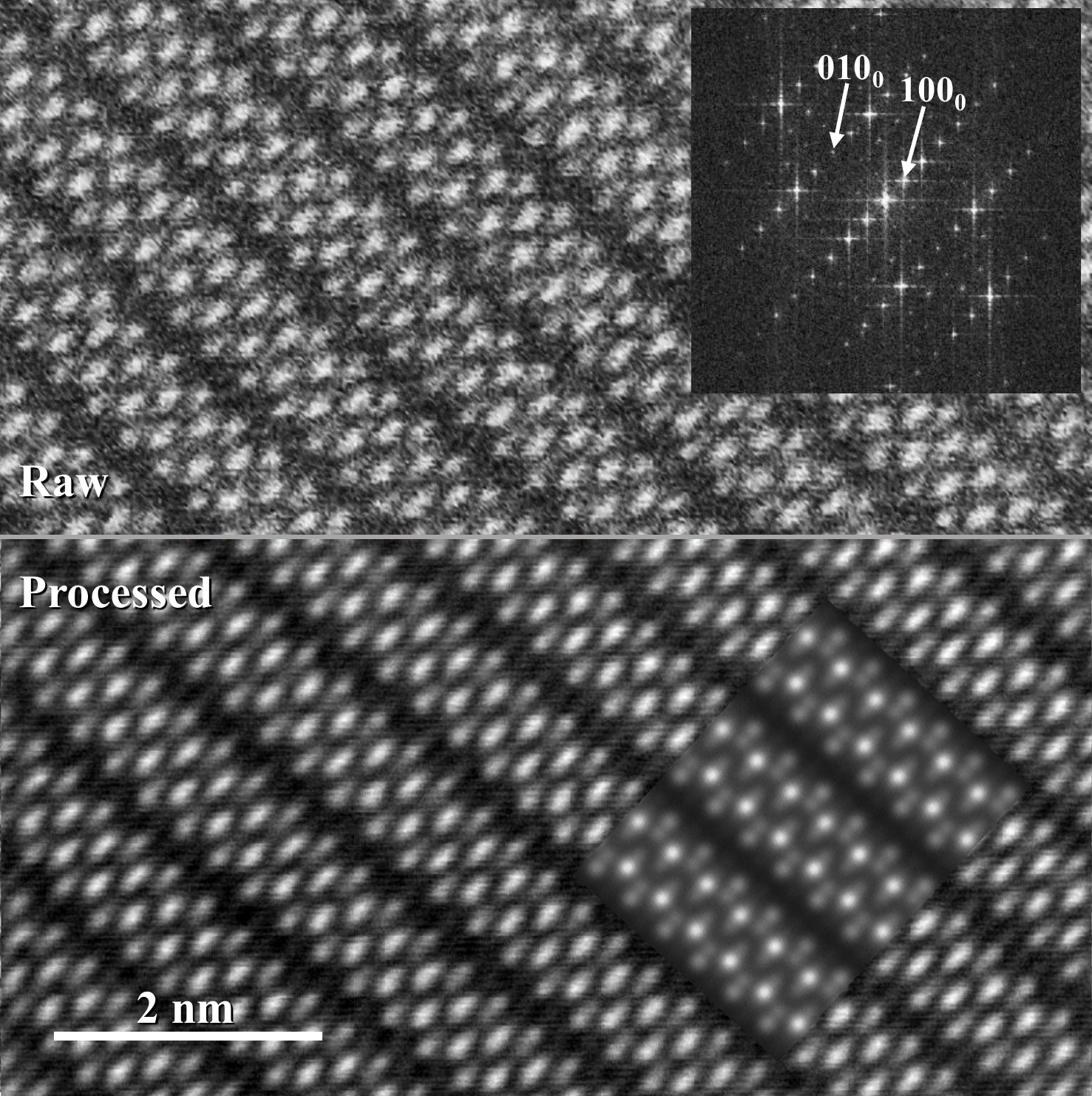}} \\
\subfloat[fig1b][]{\includegraphics[width=\linewidth]{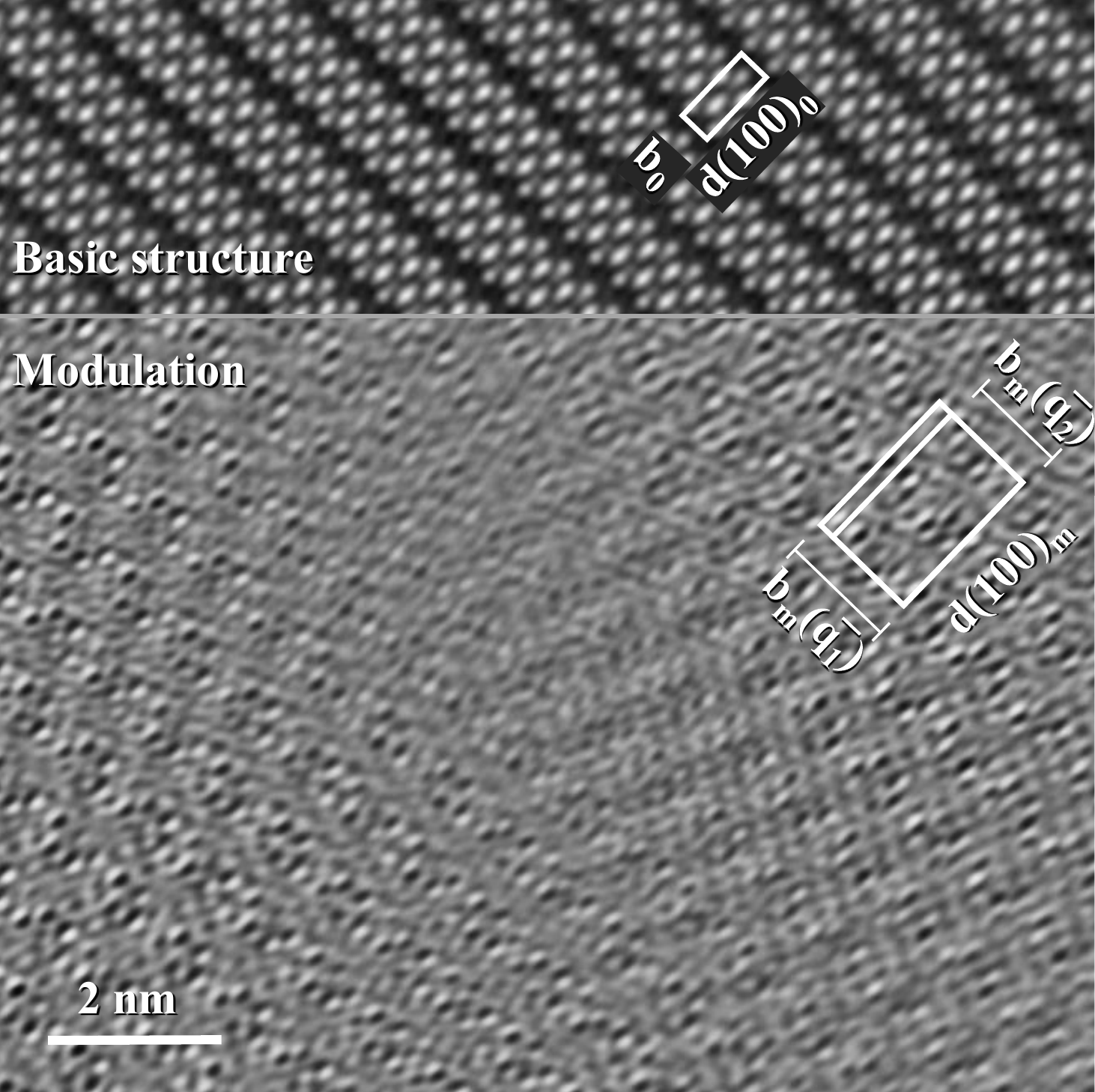}}
\caption{(a) Raw (upper region) and processed (bottom region) HAADF-STEM image of NbS\sub{3}-II, viewed along the [001] direction. The corresponding FFT image is shown in the upper right inset. The calculated HAADF-STEM image of the structural model is shown overlaid in the inset at the lower right. (b) Inverse Fourier transform images formed separately with the strong basic structure reflections (upper narrow region) and with the weak incommensurate satellites (wider bottom region). The projected unit cells of both the basic structure and the CDW modulation are outlined. Note that there is a phase shift of $\pi$ between the CDWs of adjacent columns.}
\label{fig:fig1}
\end{figure}

\textit{Atomic-resolution high-angle angular dark-field scanning transmission electron microscopy.}--Figure~\ref{fig:fig1}a shows a raw and a processed HAADF-STEM image of NbS\sub{3}-II with the beam aligned along the [001] zone axis. An inset of a simulated image based on the proposed structural model, determined by combining experimental (XRD, HR STEM and LT STM) and theoretical (ab-initio calculations) studies, is shown overlaid on the processed experimental image. A fast Fourier transform (FFT) of the experimental image is shown in the upper right inset, with only the $\vec{q_2}$ IC satellites clearly resolved in this instance. It is consistently observed that the satellites, whether present as complete doublets or as $\vec{q_2}$ singlets, appear relatively sharp only in the shown [001] zone axis patterns. By tilting the sample from the [001] toward the [100] zone around the $\vec{b_0}$ axis, the satellites in the DPs become progressively weaker and elongated to form streaks connecting the IC satellites perpendicular to the $\vec{b\sub{0}\textsuperscript{*}}$ direction. These streaks become essentially undetectable in the [100] zone axis pattern. Consequently, the HR STEM images, recorded with the beam along the NbS\sub{3}-II [100] zone axis, are generally of poor quality in comparison with those recorded along the [001] crystal zone. The streaking in the DPs arises from the presence of numerous defects, clearly detected along the $\vec{c_0}$-direction in STM images of the (100) VdW surfaces. The HR STEM images with the corresponding DPs and FFTs, recorded along the [100] zone axis, are therefore useless for structural modeling.

After FFT filtering and inverse FFT (IFFT) of the Figure~\ref{fig:fig1}a evidence of IC CDW modulation is clearly observed, as shown in Figure~\ref{fig:fig1}b. Most often only the stronger ($\vec{q_2}$) of the two satellites forming a pair is detected in the DPs and the FFTs. The second satellite ($\vec{q_1}$) is always weaker or even absent at room temperature\cite{5,9} (as in the case for the image shown here). The reason for the absence of the $\vec{q_1}$ satellites in the DPs is either
sample heating above the $\vec{q_1}$ onset temperature T\sub{1} by electron irradiation, or a variation in stoichiometry, which possibly influences the formation and onset temperature of the $\vec{q_1}$ CDW. It should be pointed out that recent linear and non-linear transport studies revealed samples showing only the T\sub{2} transition, without any evidence of the T\sub{1} transition \cite{vadim, vadim2}. In accord with the doubled modulation unit cell along the $\vec{a_0}$ direction, the IFFT image formed with satellites only (bottom region of Figure~\ref{fig:fig1}b) shows areas with strong and weak CDW modulation and with a phase shift of $\pi$ between adjacent columns. The areas with less pronounced modulation can be a result of beating interference between the slightly different modulation periods of the two CDWs. However, the modulation does not vanish completely because the $\vec{q_1}$ satellites have much weaker intensity in comparison to the $\vec{q_2}$ satellites. As a consequence, the corresponding modulation amplitudes of $\vec{q_1}$ are also much smaller and beating between the two modes cannot extinguish the modulation completely.\\

\begin{figure}[ht!]
\centering
\includegraphics[width=\linewidth]{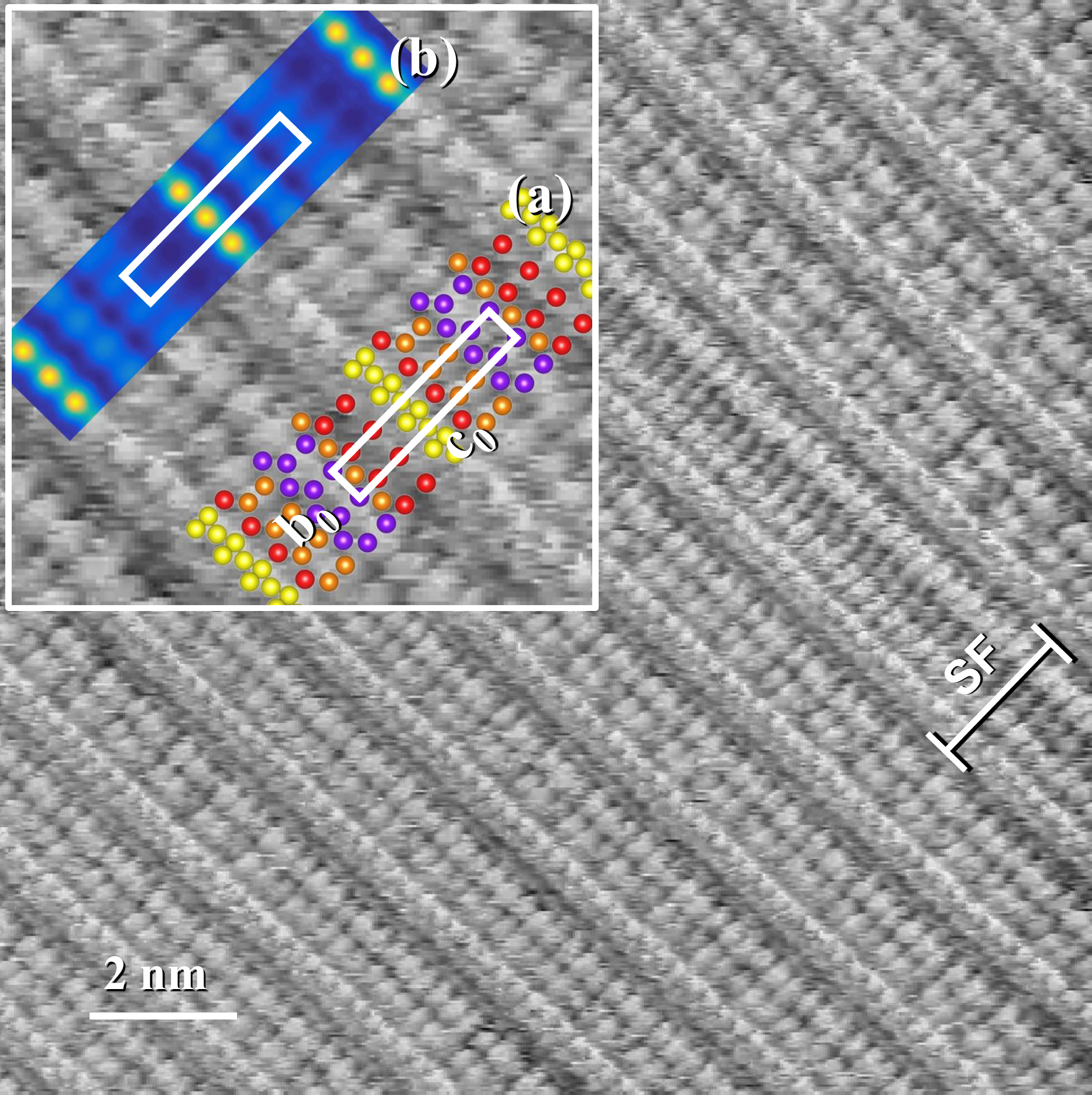}
\caption{\label{fig:fig2} 13.5 x 10 nm\textsuperscript{2} constant current STM image of the (100) Van der Waals surface of NbS\sub{3}-II, with resolved surface S columns along the $\vec{b_0}$-direction. In the enlarged inset, the basic structure unit cell is outlined and overlaid onto the projection of the calculated model structure (a) and onto the corresponding simulated STM image (b). Different types of columns are marked by the same colors as in Figure~\ref{models}. A stacking fault (SF) is indicated. In the projection of the calculated model structure, only topmost S atoms of each column are shown. Note the in-phase modulations along the S chains belonging to adjacent type-II (O) and type-IV (P) columns. Imaging parameters: $U\sub{sample bias}$ = -1.02V, $I\sub{tunneling}$ = 282 pA, $T$ = 140K. }
\end{figure}

\textit{Low-temperature scanning tunneling microscopy.}--Since high-resolution HAADF-STEM images of the [100] zone failed to give proper information for structural modeling, LT STM was employed to study the structure of the (100) VdW surface of NbS\sub{3}-II. The method is convenient in this particular case, because it largely excludes contribution from sub-surface planar defects. A LT STM image of a freshly cleaved (100) VdW surface is shown in Figure~\ref{fig:fig2}. The basic structure unit cell is overlaid in the inset onto the projection of the calculated structure and onto the corresponding simulated \footnote{STM image simulations are based on a large supercell with some vacuum, 1 x 39 x 9 k-points grid, and an energy cut-off of 250 eV.} STM image. In addition to the regular periodicity formed by four pairs of columns, three isosceles (type-I or orange-O, type–III or yellow-Y and type-IV or purple-P) and one almost equilateral (type-II or red-R), the image also reveals one longer period along the $\vec{c_0}$-direction, representing a stacking fault (labeled as SF in Figure~\ref{fig:fig2}). Similarly, shorter periods were also observed. The STM simulation based on the overlaid model structure, gives several salient features. First, the modulated O and purple P columns appear brighter than the unmodulated red R columns. Second, there is a slight contrast of the two pairs of O and P columns per unit cell. Third, the yellow Y columns are the brightest but show almost no CDW modulation. All these features are consistent between the STM modeling and experiment. Furthermore, the bright protrusions of the Y columns are out-of-phase from the bright features of the O and P columns. This is due to the arrangement of bi-capped adjacent NbS\sub{3} columns as shown in Figure~\ref{fig:fig1}. Therefore, the four different pairs of columns can be clearly distinguished. The arrangement of the columns and the symmetry restrictions of the modulation space group require that both columns forming a pair, with one of them on top and the other below the surface, are alternatively modulated by $\vec{q_1}$ and $\vec{q_2}$ modes. The amplitudes of the modes along different columns vary between very strong and almost absent, which can be a result of beating between the coupled $\vec{q_1}$ and $\vec{q_2}$ modes. This is similar to what was found in NbSe\sub{3} \cite{11,27} and indicates that the actual building units of the modulated structure are pairs of TP columns with isosceles bases, whose IC components along the $\vec{b_0}$-direction add to a COM value. It is also suggestive that the $\vec{q_1}$ modes, often absent in the electron DPs are in fact present, but are (contrary to the $\vec{q_2}$ modes) disordered along the columns above the $T\sub{1}$ onset temperature.

Another feature revealed in the LT STM image is the CDW modulation phase shift between columns of adjacent unit cells in the $\vec{c_0}$-direction. This phase shift is not fixed; it can be close to either in-phase, out-of-phase, or arbitrary. This disorder can account for the progressive weakening and streaking of the IC satellites in the electron DPs on approaching the [100] zone axis. The origin for this disorder could be the relatively large transversal distances (weaker coupling) between the CDW columns columns and/or a frequent interchanging of the related $\vec{q_1}$ and $\vec{q_2}$ modes. The STM image thus indicates that the CDW ordering is restricted to domains and extends at most over several unit cells along the $\vec{c_0}$-direction.

Finally, as observed in NbSe\sub{3}, the R columns in NbS\sub{3}-II are not modulated except for possible minor adjustments due to the modulations in the neighboring columns. In contrast to the R columns, the Y columns appear very intense, but show no obvious modulation. The origin of the strong Y columns contrast is likely in topography (since the Y columns are protruding into the VdW gaps). An enhanced charge distributed along the Y columns could also contribute. \\

\textit{Ab-initio calculation of the basic structure.}--It is known\cite{22-2,23-2} that the DFT method, if applied to layered structures like NbS\sub{3}-II, allows an unrealistic extension of the structure into the VdW gap. Consequently, all relaxations were initially performed with locked $b\sub{0}:a\sub{0}$ and $c\sub{0}:b\sub{0}$ ratios. In all cases considered, the value of $b\sub{0}$ (0.3365 nm) was chosen as half of the value determined by Rijnsdorp and Jellinek\cite{4} for NbS\sub{3}-I, while the corresponding approximate parameters $a\sub{0}$ and $c\sub{0}$ were estimated from HAADF-STEM and STM experiments. The refinements require that the angle $\beta$ be finally optimized by varying $a_0$ as a function of $\beta_0$ with all remaining parameters kept constant. From all models considered, only the accepted one was found to be in accord with both STM and HAADF-STEM images. A XRD Le Bail fitting of the powder diagrams confirmed the chosen model was in fact in accord with the lattice parameters, reported previously by Van Smaalen\cite{vansmaalen}. A second ab-initio refinement of the structural model using the transformation of Van Smaalen’s parameters ($a\sub{0}$ = 0.9654 nm, $b\sub{0}$ = 0.3346 nm, $c\sub{0}$ = 1.9855 nm, $\beta_0$ = 110.71\dg) gave the final atomic positions. The structure of the best fitting model is shown in Figure~\ref{models}. The corresponding structural data were additionally improved by a Rietveld refinement and are collected in Table~\ref{table1}. \\

\captionsetup[subfloat]{labelformat=empty}

\begin{table}[ht!]
\centering

		\subfloat[table1][]{
		\begin{tabular}{ | p{1.2cm} | p{1.2cm} | p{1.6cm} | p{1.6cm} | p{1.6cm} |}
		\hline
		\multicolumn{5}{|l|}{$a_0$ = 0.96509(8) nm, $b_0$ = 0.33459(2)  nm, $c_0$  = 1.9850(1) nm} \\
		\multicolumn{5}{|l|}{$\beta_0$ = 110.695(4)\dg, Rwp = 5.75\%, GOF = 2.68} \\
		\multicolumn{5}{|l|}{Space group $P2_1/m \hspace{1mm} (no. 11)$} \\
		\multicolumn{5}{|l|}{Nb and S atoms in $2e$ special positions} \\
		\hline
		\hline

		Atom & Column & x\sub{0} & y\sub{0} & z\sub{0} \\ \hline
		Nb 	&	Y	&	0.7005(8)	&	0.75000	&	0.5280(3)	\\	\hline
		S  	&	Y	&	0.546(2)	&	0.25000	&	0.5767(9)	\\	\hline
		S  	&	Y	&	0.913(2)	&	0.25000	&	0.5490(8)	\\	\hline
		S  	&	Y	&	0.740(2)	&	0.25000	&	0.4422(9)	\\	\hline
		Nb 	&	R	&	0.6893(9)	&	0.25000	&	0.7217(3)	\\	\hline
		S  	&	R	&	0.840(2)	&	0.75000	&	0.8071(9)	\\	\hline
		S  	&	R	&	0.828(2)	&	0.75000	&	0.6751(9)	\\	\hline
		S  	&	R	&	0.469(2)	&	0.75000	&	0.681(1)	\\	\hline
		Nb	& 	O	&	0.368(1)	&	0.75000	&	0.7876(4)	\\	\hline
		S   &	O	&	0.178(2)	&	0.25000	&	0.7946(9)	\\	\hline
		S  	&	O	&	0.204(2)	&	0.25000	&	0.697(1)	\\	\hline
		S 	&	O	&	0.579(2)	&	0.25000	&	0.8297(9)	\\	\hline
		Nb 	&	P	&	0.662(1)	&	0.25000	&	0.9675(4)	\\	\hline
		S  	&	P	&	0.835(3)	&	0.75000	&	0.056(1)	\\	\hline
		S  	&	P	&	0.842(2)	&	0.75000	&	0.9461(9)	\\	\hline
		S  	&	P	&	0.446(3)	&	0.75000	&	0.930(1)	\\	\hline

		\end{tabular}}

\caption{Structural parameters of the basic structure of NbS\sub{3}-II, obtained by ab-initio calculation and improved by the Rietveld refinement of a synchrotron x-ray powder diffraction pattern. O, R, Y and P refer to type-I, type-II, type-III and type-IV columns, respectively.}
\label{table1}
\end{table}

\begin{figure}[ht!]
\centering
\includegraphics[width=\linewidth]{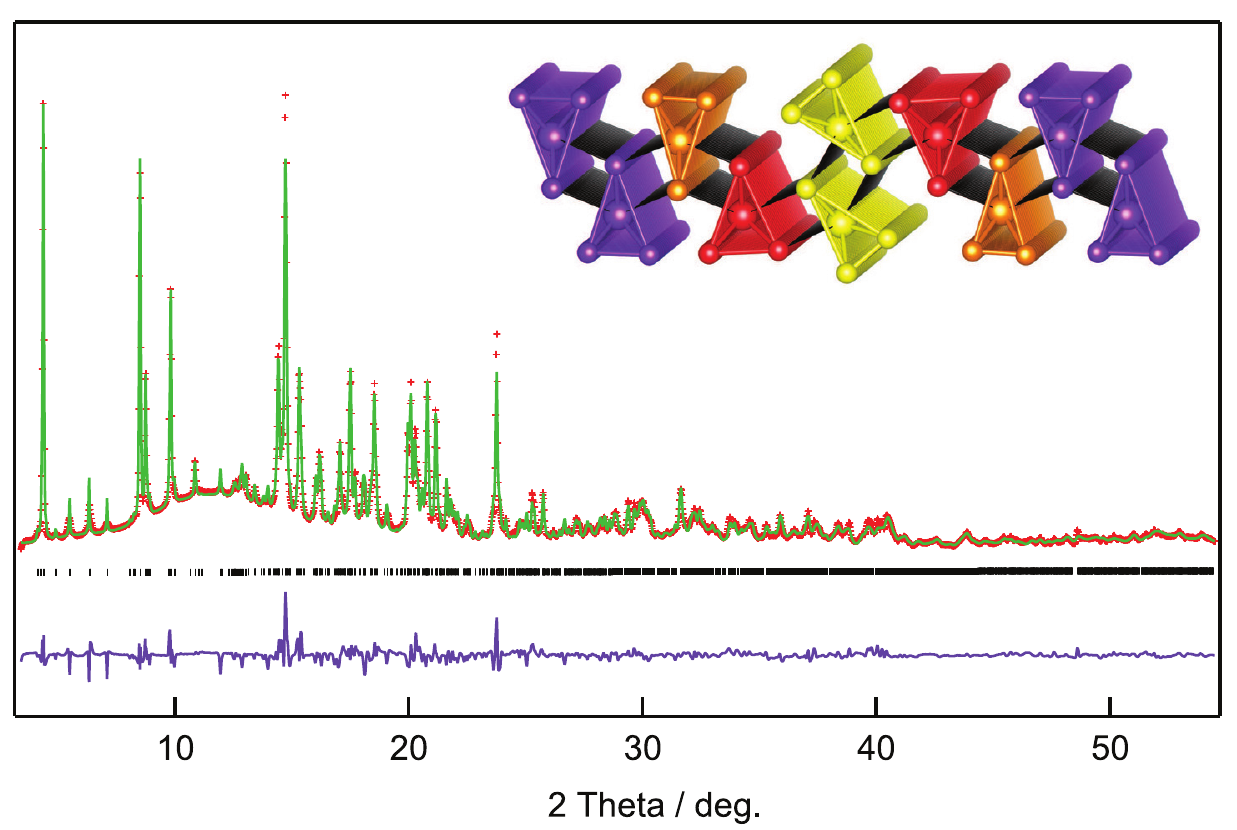}
\caption{\label{fig_xrd} Rietveld refinement of a NbS\sub{3}-II synchrotron x-ray diffraction pattern, collected at 300 K. The red crosses and the green and purple lines represent the observed pattern, the calculated profile and the difference between the observed and the calculated intensities, respectively. The Bragg peaks are shown as black tick marks. The inset shows a three-dimensional model of the structure.} 
\end{figure}

\begin{figure}[ht!]
\centering
\includegraphics[width=\linewidth]{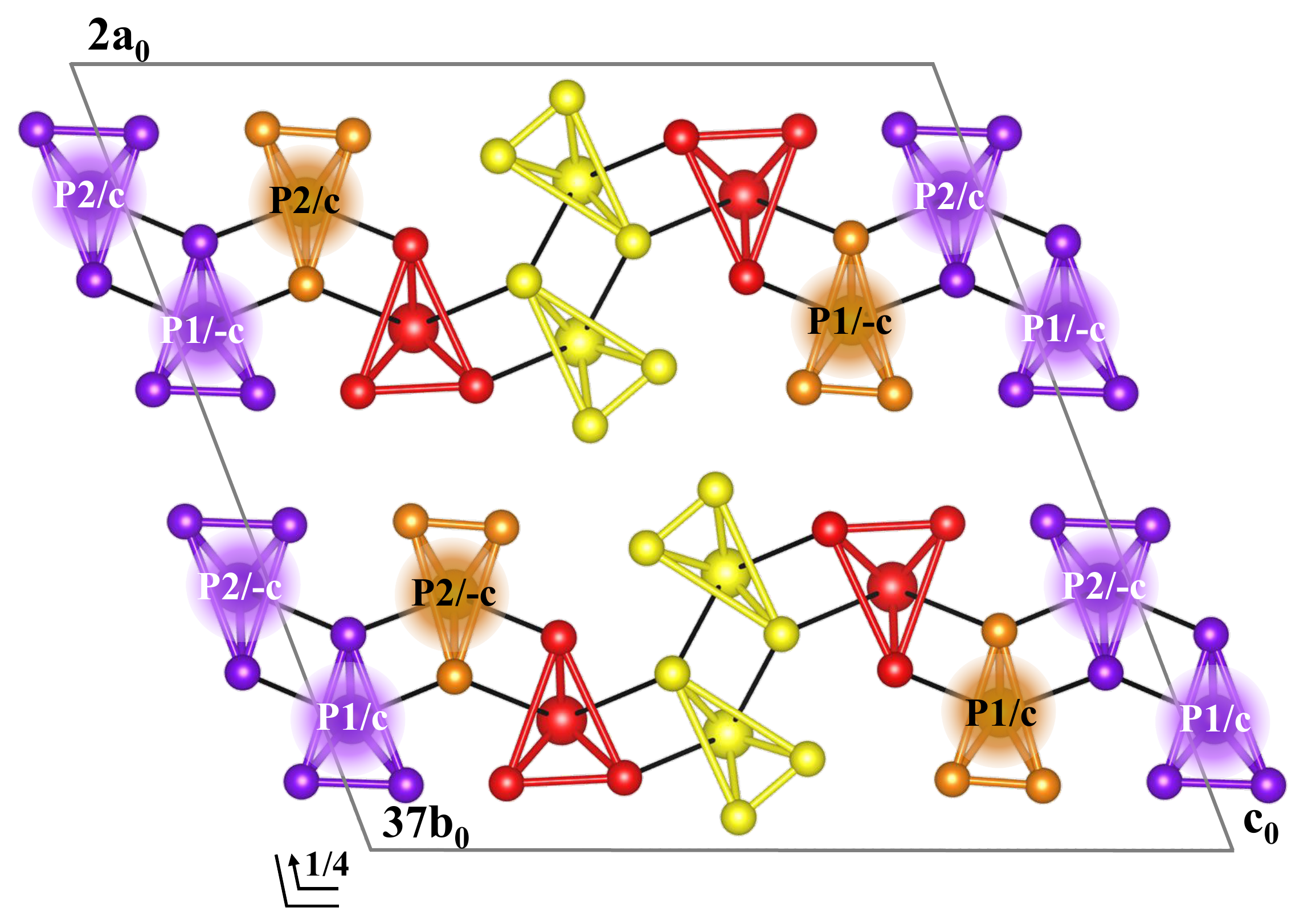}
\caption{\label{fig_modulation} The modulation unit cell of NbS\sub{3}-II with one of the two possible arrangements of the $\vec{q_1}$ and $\vec{q_2}$ modes. Nb and S atoms are shown as in Figure~\ref{models}. The modulated columns are shown as clouds with the colors of the columns, with -c and c standing for $-cos$ and $cos$ functions and P1 and P2 for the two $\vec{q}$-vectors modulating adjacent columns. The second possible arrangement is obtained by interchanging all P1 and P2 values. Symmetry operations of the modulation space group $Cm$ are added.} 
\end{figure}

\textit{Rietveld refinement of the basic structure.}--A Rietveld refinement was performed against the XRD data and is shown in Figure~\ref{fig_xrd}. The refinement started with data obtained by ab-initio calculation and resulted in an acceptable weighted profile R-factor (Rwp) of 5.75\% and a goodness of fit (GOF) factor of 2.68). The resultant lattice constants are $a\sub{0}$ = 0.96509(8) nm, $b\sub{0}$ = 0.33459(2) nm, $c\sub{0}$ = 1.9850(1) nm, $\beta_0$ = 110.695(4)\dg, $Z$ = 8, space group $P2_1/m \hspace{1mm} (no. 11)$ with all Nb and S atoms in special $2e$ positions. The refined structural parameters of the Figure~\ref{models} model are given in Table~\ref{table1}.\\

\textit{Charge density wave modulated structure.}--Because of the limited computational power, ab-initio calculations of the modulated structures, considered as LP COM with hundreds of atoms per unit cell, were not feasible to perform. Instead, DPs of different models, based on the determined basic structure, were calculated. To include both IC $\vec{q}$-vectors, the basic structure unit cell had to be accordingly enlarged along the monoclinic $\vec{b_0}$-axis (to coincide within acceptable errors with multiples of both CDW periodicities) and doubled along the $\vec{a_0}$-axis (to comply with the extinction rules of the modulation determined from electron DPs). Three possible combinations of $q_{1b} / nb_{0}$ and $q_{2b} / nb_{0}$ were considered; they are ($2a\sub{0}$, $nb\sub{0}$,  $c\sub{0}$, $\beta$) with $n$ = 17, 36 and 37, correspond to $q_{1b}$ = 5/17, 11/36 and 11/37 and $q_{2b}$ = 6/17, 13/36 and 13/37, respectively. The first set of ratios was suggested earlier\cite{5} as the simplest possibility, requiring the smallest enlargement of the modulated structure unit cell. The second pair fits exactly to a COM value (11/36 + 13/36 = 2/3), while the third case considered ($\vec{q_{1b}}$ = 11/37 = 0.2973 and $\vec{q_{2b}}$ = 13/37 = 0.3514) gives the best fit with the experimentally determined IC components of the $\vec{q}$-vectors, i.e. 0.298${b_0^*}$ and 0.352${b_0^*}$ \cite{2}. The enlarged cells contain in the three cases altogether 1088 (272 Nb and 816 S), 2304 (576 Nb and 1728 S) and 2368 (592 Nb and 1776 S) atoms, respectively. These combinations were chosen to verify the positions and the appearance/disappearance of the weak IC satellites in the calculated DPs, particularly those recorded along the most important [100] and [001] zone axes.

Since the NbS\sub{3}-II basic structure is closely related to that of NbSe\sub{3} and because pairs of $\vec{q}$-vectors detected in both compounds add approximately to COM values, it is expected that the CDW ordering in both compounds obeys similar restrictions\cite{11}. According to these restrictions, the two $\vec{q}$-vectors should also be formed only pairwise along two adjacent, the same or very similar columns (in NbSe\sub{3} of the same type). The resulting CDW LP units evolve into temperature dependent modulation structures due to ordered or disordered arrangements of these units. This pairwise formation is a common feature of the experimentally determined fact that the IC components of both $\vec{q}$-vectors add within experimental error into a COM value (in NbSe\sub{3} $q_{1b} + q_{2b} = 1/2$ and in NbS\sub{3}-II $q_{1b} + q_{2b} \approx 2/3$). Consequently, the R columns will not be modulated, not only because they are more "equilateral-like", but rather because they are in both compounds spatially separated. The two $\vec{q}$-vectors in NbS\sub{3}-II can only occupy certain pairs of columns, the O, the Y and the P pairs. However the O columns can only be modulated when the intervening P columns are also modulated. These pairs of CDWs must be ordered in accord with the extinction rules of the modulation space group $C2/m \hspace{1mm} (no. 12)$, determined on the basis of electron diffraction experiments\cite{10}, or some other space group obeying the same extinction rules, i.e. $C2 \hspace{1mm} (no. 5)$ and $Cm \hspace{1mm} (no. 8)$. Among these three space groups $Cm$ is in the best consistency with the LT STM images. It shows that the two surface modes along the P and O columns belong to the same $\vec{q}$-vector (either $\vec{q_1}$ or $\vec{q_2}$), ordered in-phase with each other.

Only half of all Nb and S atoms (i.e. 8 columns with 17, 36 or 37 Nb and three times as many S atoms each, dependent on the $q_{1b}/ nb_{0}$ and $q_{2b} / nb_{0}$ values chosen) will be modulated by the $\vec{q_1}$ and $\vec{q_2}$ modes, while all remaining Nb and S atoms in the R and Y columns will not be displaced from their original positions. Since the intensities of all IC satellites in DPs are very weak, the corresponding modulation amplitudes must also be very small. Although these intensities cannot be accurately determined from electron DPs, they do indicate that the actual modulation is rather confined to the lighter S instead of the heavier Nb atoms. This is also supported by the fact, clearly demonstrated during modeling, that all possible modulation space groups can be strictly obeyed only for transversal displacements, perpendicular to the $\vec{b_0}$-direction. Thus, the only reasonable solution is a "breathing mode", formed of S atoms, leaving all Nb atoms in their equilibrium positions. The corresponding modulation amplitudes are thus confined to displacements in the (010) planes, while the phase-shifts between the columns are restricted by the symmetry of the modulation space group. One of the two possible arrangements of the two modes is shown in Figure~\ref{fig_modulation}. The second is obtained by interchanging the $\vec{q_1}$ and $\vec{q_2}$ modes, i.e. by simply interchanging all P1 and P2 values, which will leave all symmetry operations of the space group $Cm$ unchanged. Different modulated columns are shown in Figure~\ref{fig_modulation} with the same colors as the corresponding columns in Figure~\ref{models}. After the required enlargement of the basic structure unit cell the (x\sub{0}, y\sub{0}, z\sub{0}) coordinates of the S atoms belonging to the O and the P columns are replaced by (x, y, z) coordinates, in accord with the modulation scheme in Table~\ref{table2}. Again, to obtain the second possible arrangement, all $\vec{q_1}$ and $\vec{q_2}$ modes (i.e. all P1 and P2 values) in Table~\ref{table3} should be simply interchanged.

\begin{table}[ht!]
\centering
	
		\subfloat[table1b][]{
		\begin{tabular}{ | p{1.3cm} | p{1.3cm} | p{1.3cm} | p{1.3cm} | p{1.3cm} |}
		\hline
		\multicolumn{5}{|l|}{$a_m$ = $2a_0$, $b_m$ = $b_0/q_{jb}$ ($j$=1,2), $c_m$ = $c_0$, $\beta_m$ = $\beta_0$}\\
		\multicolumn{5}{|l|}{Space group $Cm \hspace{1mm} (no. 8)$} \\
		\multicolumn{5}{|l|}{Both CDWs related as $2a$ special positions} \\
		\hline
		\hline
		CDW & column &x\sub{m} & y\sub{m} & z\sub{m} \\ \hline
		$\vec{q_1}$ & O & 0.17485 & 0.00000 & 0.79057 \\ \hline
		$\vec{q_1}$ & P	& 0.16619 & 0.00000 & 0.03656 \\ \hline
		$\vec{q_2}$ & O & 0.32515 & 0.50000 & 0.20943 \\ \hline
		$\vec{q_2}$ & P	& 0.33381 & 0.50000 & 0.96344 \\ \hline
		\end{tabular}}
		
\caption{The structural parameters of the $\vec{q_1}$  and $\vec{q_2}$ CDW modulations for one of two possible arrangements. O and P refer to type-I and type-IV columns, respectively. The alternate arrangement is obtained by interchanging $\vec{q_1}$ and $\vec{q_2}$ on O and P columns.}
\label{table2}
\end{table}

\begin{table*}[ht!]
\centering
		
		\subfloat[table1c][]{
		\begin{tabular}{ | p{0.8cm} | p{1.2cm} | p{6.3cm} | p{1.9cm} | p{6.3cm} |}
		\hline
		\multicolumn{5}{|l|}{$a$ = $2a_0$, $b$ = $nb_0$ ($n$=36,37), $c$ = $c_0$, $\beta$ = $\beta_0$, Space group $P1 \hspace{1mm} (no. 1)$}\\
		\multicolumn{5}{|l|}{All atoms in general $1a$ positions; all Nb atoms and the S atoms of type-II (R) and type-III (Y) columns in unmodulated positions given} \\		
		\multicolumn{5}{|l|}{in Table~\ref{table1} and the remaining S atoms of type-I (O) and type-IV (P) columns displaced according to the following modulation scheme:} \\
		\hline
		\hline
		Atom & Column & x & y & z \\ \hline
		S & O(1/-c) &0.089100$\cdot$(-A(-0.69974)$\cdot$cos[(m/n+1/(4n))$\cdot$P1]) & [m/n+1/(4n)] & 0.79469$\cdot$(-A(0.71440)$\cdot$cos[(m/n+1/(4n))$\cdot$P1])\\ \hline
S & O(1/-c) &0.10210$\cdot$(-A(-0.82646)$\cdot$cos[(m/n+1/(4n))$\cdot$P1]) & [m/n+1/(4n)] & 0.69710$\cdot$(-A(0.56299)$\cdot$cos[(m/n+1/(4n))$\cdot$P1])\\ \hline
S & O(1/-c) &0.28960$\cdot$(-A(-0.19717)$\cdot$cos[(m/n+1/(4n))$\cdot$P1]) & [m/n+1/(4n)] & 0.82979$\cdot$(-A(0.98037)$\cdot$cos[(m/n+1/(4n))$\cdot$P1])\\ \hline
S & P(2/+c) &0.41765$\cdot$(+A(0.10895)$\cdot$cos[(m/n+3/(4n))$\cdot$P2]) & [m/n+3/(4n)] & 0.056100$\cdot$(+A(-0.99405)$\cdot$cos[(m/n+3/(4n))$\cdot$P2])\\ \hline
S & P(2/+c) &0.42110$\cdot$(+A(0.32224)$\cdot$cos[(m/n+3/(4n))$\cdot$P2]) & [m/n+3/(4n)] & 0.94619$\cdot$(+A(0.94666)$\cdot$cos[(m/n+3/(4n))$\cdot$P2])\\ \hline
S & P(2/+c) &0.22315$\cdot$(+A(-0.50386)$\cdot$cos[(m/n+3/(4n))$\cdot$P2]) & [m/n+3/(4n)] & 0.93010$\cdot$(+A(0.86379)$\cdot$cos[(m/n+3/(4n))$\cdot$P2])\\ \hline
S & P(1/-c) &0.27685$\cdot$(-A(0.12823)$\cdot$cos[(m/n+1/(4n))$\cdot$P1]) & [m/n+1/(4n)] & 0.069900$\cdot$(-A(-0.99174)$\cdot$cos[(m/n+1/(4n))$\cdot$P1])\\ \hline
S & P(1/-c) &0.078900$\cdot$(-A(-0.14184)$\cdot$cos[(m/n+1/(4n))$\cdot$P1]) & [m/n+1/(4n)] & 0.053810$\cdot$(-A(-0.98989)$\cdot$cos[(m/n+1/(4n))$\cdot$P1])\\ \hline
S & P(1/-c) &0.082350$\cdot$(-A(-0.54548)$\cdot$cos[(m/n+1/(4n))$\cdot$P1]) & [m/n+1/(4n)] & 0.94390$\cdot$(-A(0.83812)$\cdot$cos[(m/n+1/(4n))$\cdot$P1])\\ \hline
S & O(2/+c) &0.21040$\cdot$(+A(-0.14961)$\cdot$cos[(m/n+3/(4n))$\cdot$P2]) & [m/n+3/(4n)] & 0.17021$\cdot$(+A(-0.98874)$\cdot$cos[(m/n+3/(4n))$\cdot$P2])\\ \hline
S & O(2/+c) &0.39790$\cdot$(+A(0.10008)$\cdot$cos[(m/n+3/(4n))$\cdot$P2]) & [m/n+3/(4n)] & 0.30290$\cdot$(+A(-0.99498)$\cdot$cos[(m/n+3/(4n))$\cdot$P2])\\ \hline
S & O(2/+c) &0.41090$\cdot$(+A(0.10419)$\cdot$cos[(m/n+3/(4n))$\cdot$P2]) & [m/n+3/(4n)] & 0.20531$\cdot$(+A(-0.99456)$\cdot$cos[(m/n+3/(4n))$\cdot$P2])\\ \hline
S & O(1/+c) &0.58910$\cdot$(+A(0.48273)$\cdot$cos[(m/n+1/(4n))$\cdot$P1]) & [m/n+1/(4n)] & 0.79469$\cdot$(+A(0.87577)$\cdot$cos[(m/n+1/(4n))$\cdot$P1])\\ \hline
S & O(1/+c) &0.60210$\cdot$(+A(0.54599)$\cdot$cos[(m/n+1/(4n))$\cdot$P1]) & [m/n+1/(4n)] & 0.69710$\cdot$(+A(0.83779)$\cdot$cos[(m/n+1/(4n))$\cdot$P1])\\ \hline
S & O(1/+c) &0.78960$\cdot$(+A(0.61425)$\cdot$cos[(m/n+1/(4n))$\cdot$P1]) & [m/n+1/(4n)] & 0.82979$\cdot$(+A(0.78911)$\cdot$cos[(m/n+1/(4n))$\cdot$P1])\\ \hline
S & P(2/-c) &0.91765$\cdot$(-A(0.96789)$\cdot$cos[(m/n+3/(4n))$\cdot$P2]) & [m/n+3/(4n)] & 0.056100$\cdot$(-A(-0.25136)$\cdot$cos[(m/n+3/(4n))$\cdot$P2])\\ \hline
S & P(2/-c) &0.92110$\cdot$(-A(0.63622)$\cdot$cos[(m/n+3/(4n))$\cdot$P2]) & [m/n+3/(4n)] & 0.94619$\cdot$(-A(0.77151)$\cdot$cos[(m/n+3/(4n))$\cdot$P2])\\ \hline
S & P(2/-c) &0.72315$\cdot$(-A(0.49345)$\cdot$cos[(m/n+3/(4n))$\cdot$P2]) & [m/n+3/(4n)] & 0.93010$\cdot$(-A(0.86977)$\cdot$cos[(m/n+3/(4n))$\cdot$P2])\\ \hline
S & P(1/+c) &0.77685$\cdot$(+A(0.94814)$\cdot$cos[(m/n+1/(4n))$\cdot$P1]) & [m/n+1/(4n)] & 0.069900$\cdot$(+A(-0.31786)$\cdot$cos[(m/n+1/(4n))$\cdot$P1])\\ \hline
S & P(1/+c) &0.57890$\cdot$(+A(0.88361)$\cdot$cos[(m/n+1/(4n))$\cdot$P1]) & [m/n+1/(4n)] & 0.053810$\cdot$(+A(-0.46822)$\cdot$cos[(m/n+1/(4n))$\cdot$P1])\\ \hline
S & P(1/+c) &0.58235$\cdot$(+A(0.53999)$\cdot$cos[(m/n+1/(4n))$\cdot$P1]) & [m/n+1/(4n)] & 0.94390$\cdot$(+A(0.84167)$\cdot$cos[(m/n+1/(4n))$\cdot$P1])\\ \hline
S & O(2/-c) &0.71040$\cdot$(-A(0.88057)$\cdot$cos[(m/n+3/(4n))$\cdot$P2]) & [m/n+3/(4n)] & 0.17021$\cdot$(-A(-0.47392)$\cdot$cos[(m/n+3/(4n))$\cdot$P2])\\ \hline
S & O(2/-c) &0.89790$\cdot$(-A(0.97541)$\cdot$cos[(m/n+3/(4n))$\cdot$P2]) & [m/n+3/(4n)] & 0.30290$\cdot$(-A(-0.22041)$\cdot$cos[(m/n+3/(4n))$\cdot$P2])\\ \hline
S & O(2/-c) &0.91090$\cdot$(-A(0.94376)$\cdot$cos[(m/n+3/(4n))$\cdot$P2]) & [m/n+3/(4n)] & 0.20531$\cdot$(-A(-0.33062)$\cdot$cos[(m/n+3/(4n))$\cdot$P2])\\ \hline
		\end{tabular}}

\caption{The structural parameters of the NbS\sub{3}-II complete modulated structure for the two cases considered ($n$ = 36, 37 and with $m$ ranging from 0 to $n$-1). Only one of two possible arrangements is given. The second possibility is obtained by interchanging all P1 and P2 values. $P1=11\cdot2\mathit{\Pi}, 11\cdot2\mathit{\Pi}$ and $P2=13\cdot2\mathit{\Pi}, 13\cdot2\mathit{\Pi}$ for the two values of $n$, respectively. P1 and P2 stay for $\vec{q_1}$ and $\vec{q_2}$ modes, O and P refer to type-I and type-IV columns, and the subscripts +c and –c for +cos and –cos functions, respectively.}
\label{table3}
\end{table*}

\begin{figure}[ht!]
\centering
\subfloat[]{\includegraphics[height=8cm]{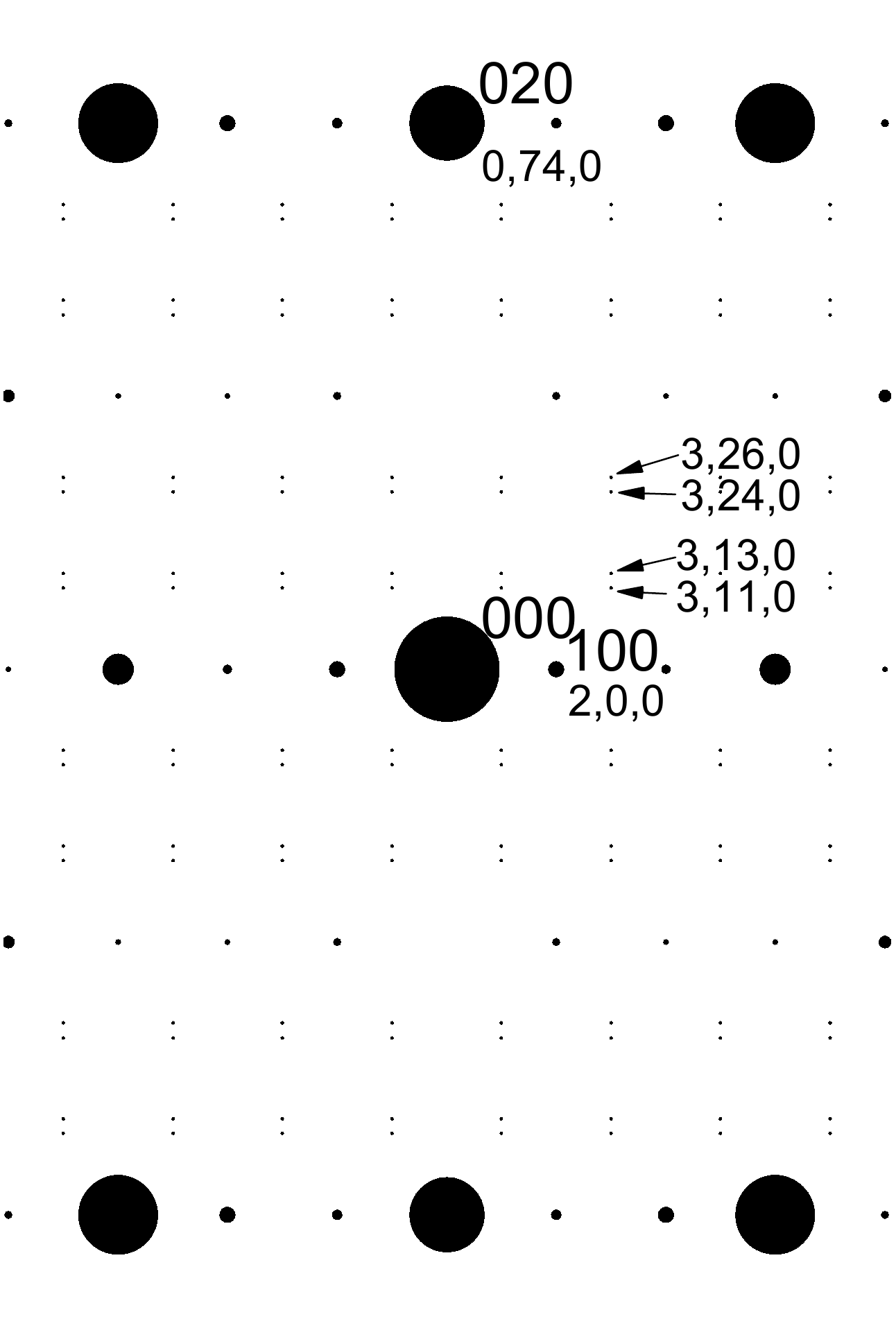}}\hfill
\subfloat[]{\includegraphics[height=8cm]{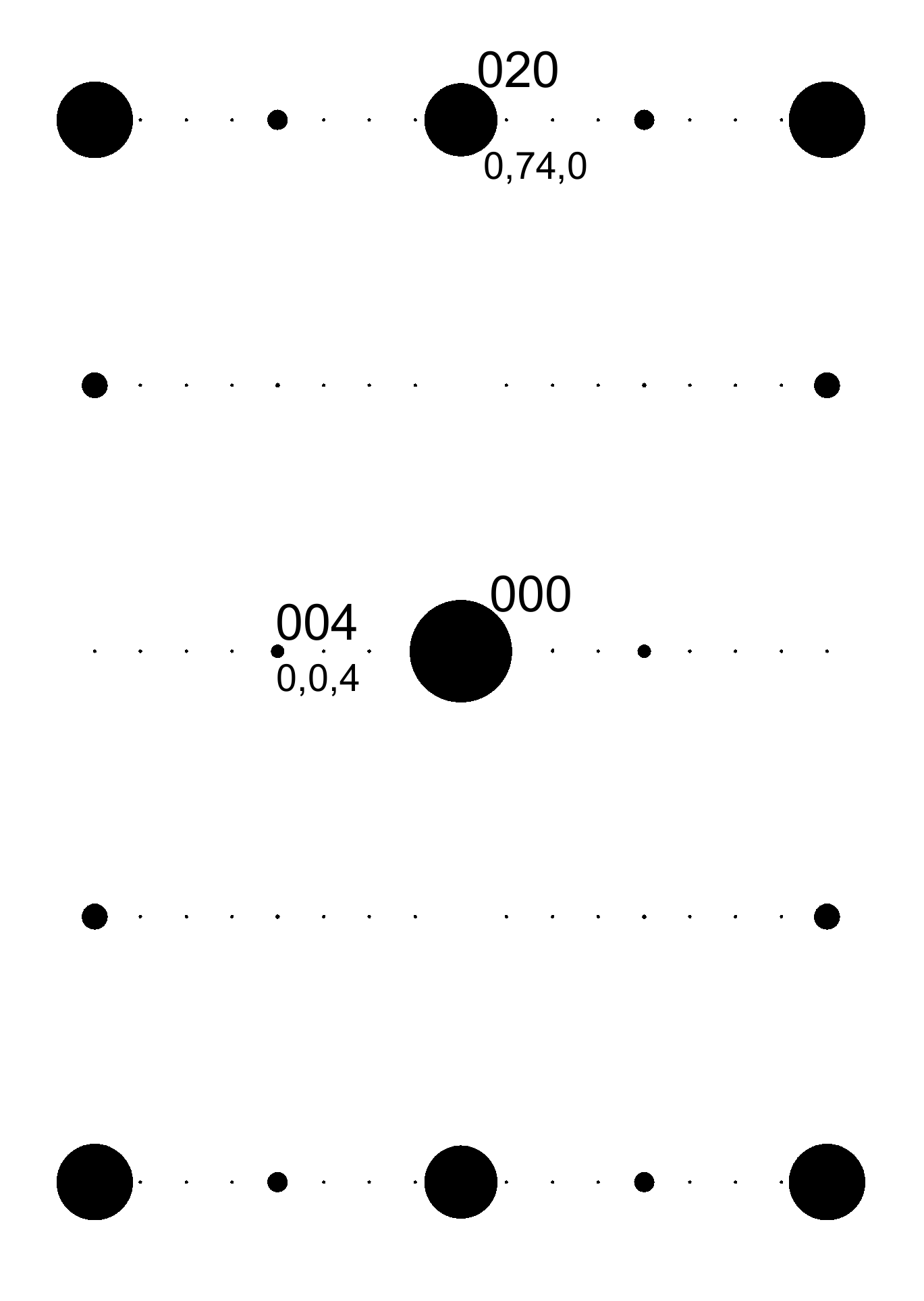}}
\caption{\label{fig_diffraction}Simulated diffraction patterns of NbS\sub{3}-II with the electron beam along the [001] (a) and [100] (b) zone axes for the proposed model with the two $\vec{q}$ -vectors corresponding to 11 ($\vec{q_1}$) and 13 ($\vec{q_2}$) periods per $37b\sub{0}$. Large and small indices refer to the basic and the modulated structures.}
\end{figure}

The simulated DPs show that from the three sets of ratios considered, only the two with 11 and 13 modulation periods per $36b\sub{0}$ and $37b\sub{0}$ give the correct results, which are almost identical. They differ only in how accurate the IC satellites are positioned with regard to the strong basic structure reflections. Best fits with the experimentally observed DPs were obtained for a relatively small modulation amplitude $A$. With the breathing mode, performed by transversely displacing the S atoms along the O ($\vec{q_1}$) and the P columns ($\vec{q_2}$), or vice versa, and with the phase shifts given in Table~\ref{table3} and in Figure~\ref{fig_modulation}, only the allowed IC satellites actually appear in all low-index zones. The calculated DPs along the two most important zones, [001] and [100], are shown in Figure~\ref{fig_diffraction} for the case with $n$ = 37. The modulation amplitudes are taken as $A$ = 0.0033 in units of $2a_0$ and $b_0$ and multiplied by factors, dependent on the direction of the displacements with respect to the $\vec{a_0}$ and $\vec{c_0}$ axes. The satellites are drawn proportional to the logarithms of the calculated intensities for all intensities over $0.05$ of the intensity of the strongest 020 basic structure reflection, or with this fixed minimum intensity for all other weaker satellites. In addition to the extinction rules, required by the modulation space group, the intensities of all satellites but the allowed ones, appear below the accuracy of the numerical calculation (bellow $10^{-7}$ of the intensity of the 020 reflection). All satellites, allowed by the modulation space group and not present in the calculated DPs, appear in the experimental electron DPs as a result of dynamical scattering. In accord with experiments\cite{6}, the sharpest and strongest pairs of IC satellites appear in electron DPs along the [001] zone axis and get weaker on rotating the crystal toward the [100] zone.  

\captionsetup[subfloat]{labelformat=parens}

Like in the case of NbSe\sub{3} the two CDWs in NbS\sub{3}-II are not equally stable at elevated temperatures. While the $\vec{q_2}$ satellites remain observed in electron diffraction experiments during repeated cycling to well above 400K, the temperature of $\vec{q_1}$ disappearance is only slightly above room temperature at about 360 K\cite{6}. However, the ordering of the two modes in NbS\sub{3}-II is different from the one in NbSe\sub{3}\cite{10}, where the two modes form different modulation unit cells, one primitive and the other centered. As a result the corresponding satellites in the electron DPs appear separated. Contrary, in case of NbS\sub{3}-II both modes form similar centered modulation cells, whose IC satellites in the DPs appear as close pairs and obey the same extinction rules.

In analogy with NbSe\sub{3} and in accord with the assumption that adjacent pairs of columns are alternatively modulated by $\vec{q_1}$ and $\vec{q_2}$, the two outer O and the two inner P column of the group of four will be modulated by the two CDWs as shown in Figure~\ref{fig_modulation}, or vice versa. Since both modes are IC with respect to the basic structure and to one another, they will also be ordered or disordered independently. The $\vec{q_1}$ satellites appear as a rule at all temperatures much weaker than the $\vec{q_2}$ ones, indicating that the corresponding modulation amplitudes are smaller and the transversal ordering of $\vec{q_1}$ less stable. The $\vec{q_1}$ mode may thus remain present above the order/disorder transition at T\sub{1} but give no detectable contribution to the DPs.

In contrast to NbSe\textsubscript{3} \cite{12}, in NbS\textsubscript{3}-II an odd number of the $\vec{q_1}$ and $\vec{q_2}$ modes corresponds equally well to either an even or an odd number of basic structure periods $n\vec{b_0}$. This is related to the extinction rules of the modulation space group $Cm$. The hk0 satellites of the LP modulated structure, present in the [001] zone, are all allowed in the space group $P1$, but the appearance of the satellites must also comply with the convolution determined by the modulation space group, which requires that h\sub{m}+k\sub{m} = even. Both modes, with only the S atoms modulated transversally, are presumed to be harmonic. Thus, only first order satellites to the strongest basic structure reflections will actually appear in the DPs. In the [100] zone, the first-order pairs of satellites to the basic structure reflections will be forbidden, while the second-order ones, though allowed, will be too weak to appear. As a result, in practice no satellites will be detected in this particular zone, in agreement with the calculated and also the experimental DPs. However, weak second order satellites may appear in case the modulation is not exactly harmonic, or if the breathing mode of the S atoms is accompanied by a weak modulation of the Nb atoms. Since Nb atoms are arranged along central chains, running in the $\vec{b_0}$-direction, they will perform a longitudinal modulation by stretching and shortening the average Nb-Nb distances.\\

\textit{Discussion.}--The space groups of the basic structure and the modulation of NbS\sub{3}-II were determined previously \cite{10}, without the knowledge of the exact arrangement of atoms in the basic structure. The current ab-initio calculations and the Rietvelt XRD refinement show that all four pairs of columns are different, with three of the four being clearly isosceles and the last more "equilateral-like". In analogy to NbSe\sub{3}, the two columns of the R pairs are spatially separated. All structural models of the basic structure considered here belonged to the space group $P2_1/m$. From different possible solutions with 2+2 and 3+1 symmetry related pairs of the four columns only the proposed model was found to be consistent with the available experimental evidence and with the ab-initio calculations. This basic structure of NbS\sub{3}-II differs from that of NbSe\sub{3} practically only in the presence of an additional P pair of isosceles columns, lying between the slightly different O columns. 

The unit cell of the modulation is clearly centered and doubled along the $\vec{a_0}$-direction. This is in accord with the space group $C2/m$, determined previously on the basis of electron diffraction experiments only\cite{10}.  However, the same extinction rules are obeyed by two further space groups, $C2$ and $Cm$. The additional STM experiments show, that the actual symmetry of the modulation is in fact reduced from $C2/m$ into $Cm$. This second space group is thus the only one of the three in accord with both, electron diffraction and STM experiments. In particular, the STM images reveal that the two surface modes, confined to the top P and O columns, indeed belong to the same mode, in-phase with one another. Thus, both modes form in the strongly bonded S-Nb-S sandwiches layers with one of the two modes on top and the other below the surface. The two modes could easily interchange and be the origin of the twinkling domains first observed in selected-area dark-field TEM images of NbSe\sub{3} \cite{fung}. It could also be a possible sliding mechanism of the CDWs \cite{27} in 1D compounds with two IC CDWs formed along adjacent pairs of columns.

The two $\vec{q}$-vectors, though originally given with non-zero x-components, are in fact directed along the columns, with only their IC y-components different from zero and with both modes ordered within a centered cell. It should be pointed out that the two IC satellites forming $\vec{q_2}$ and $\vec{q_1}$ pairs in the electron DPs appear as first- and second-order reflections. However, the extinction rules of the modulation space group contradict such a possibility. There is no space group, which would allow higher-order satellites only along lines midway between the strong basic structure reflections. Consequently, both IC satellites of a pair have to be attributed to two different, but related CDWs. This is in accord with the STM and HAADF-STEM images, but also with the suggestion that CDW sliding is a result of an easy switching between the two CDWs with slightly different IC components \cite{27}.

Regardless of whether both ($\vec{q_1}$ and $\vec{q_2}$) or only one ($\vec{q_2}$) of the IC satellites are present in the electron DPs, they are as a rule detected as weak, but relatively sharp reflections only in DPs recorded along the [001] zone axis \cite{6,10}. By rotating the crystal around the $\vec{b_0}$-axis away from the [001] zone, the satellites become weaker and more elongated, finally becoming undetectable at RT in the [100] zone axis orientation. This is in accord with the calculated DPs, where the intensities of the IC satellites become negligible as compared to those in the [001] zone. The atypical periods, often observed in LT STM images of the (100) VdW planes, in
addition to the regular periodicity with eight columns per $c_0$, are supposed to come from intersections of the surface plane with planar faults. Since the VdW gaps between the S-Nb-S sandwiches are corrugated and the columns should remain equidistant across the gaps, the regions must represent intersections of the (100) plane with stacking faults, parallel to the (001) plane.

Unlike most other MX\sub{3} compounds, NbS\sub{3}-II is formed (in addition to the spatially separated pair of the "equilateral-like" columns) of three isosceles pairs of columns. This may be related to the appearance of a third $\vec{q}$-vector below 150 K \cite{15}. This additional CDW can only be formed along the remaining Y columns. A pair of equal, but out-of-phase modes along the two Y columns, which are frequently interchanged, would average out any contribution to the DP and diminishing apparent CDWs along the Y columns. Such a disorder is possible, particularly if the nonstoichiometry in NbS\sub{3}-II results in numerous point defects, present along the Y columns. It was shown by STM \cite{26} that in the corrugated dichalcogenide $\beta$-NbTe\sub{2}, Te vacancies appear only along the topmost Te columns. The difference between corrugated di- and tri-chalcogenides is in that regard only in the orientation of the TPs forming the corrugated columns. Thus, the numerous vacancies may indeed regulate the composition of the single crystals in nonstoichiometric compounds like NbS\sub{3}-II. As in other CDW compounds (e.g. TaS\sub{2}), the defects will locally suppress the CDWs and prevent their long-range ordering. If that was also the case in NbS\sub{3}-II, it may result in a small, but still detectable CDW current under an external electric field. However, the only structural evidence that an additional short-range charge modulation may be present below 150 K along the Y columns would be an enhanced contrast of these columns in the STM images of the (100) VdW surfaces. However, the Y columns are protruding into the VdW gaps and it is practically impossible to distinguish a possible charge contribution from the topological one. In addition, the Y columns forming symmetry-related pairs are positioned almost exactly on top of each other, which further complicates any overlapped contribution of both to the STM image. Such a scenario may also explain why CDWs are not formed preferably along the isosceles Y columns (like is the case in NbSe\sub{3}), but rather along the two pairs of slightly wider O and P columns.\\

\textit{Conclusions.}--The present study offers a better understanding of the structural properties of NbS\sub{3}-II and of the entire family of MX\sub{3} compounds in general.\\
 -	The basic structure of NbS\sub{3}-II is determined by combining HAADF-STEM, STM and XRD experiments with ab-initio calculations.\\
-	It is confirmed that the basic structure unit cell of NbS\sub{3}-II is closely related to the one of NbSe\sub{3}, with only an additional isosceles pair of symmetry related TP columns.\\
-	Only half of the columns are modulated by the two CDWs, whose IC components along the columns add approximately to a COM value.\\
-	The ordering of the two IC CDWs is explained by considering them as a LP COM superstructure of the basic NbS\sub{3}-II structure.\\

\begin{acknowledgments}
The work was supported by Slovenian Research Agency (ARRS) (under the Slovenia-Russia bilateral projects (BI-RU/14-15-043 and BI-RU/16-18-048) (EZ, HJPvM, MvM, SŠ, ET, AP), by the Russian Academy of Sciences (RFBR grants 16-02-01095, 17-02-01343) (VYaP, SGZ, VFN, SVZ-Z), by the Ministry of Science and Technology of Taiwan (103-2923-M-002-003-MY3, 103-2112-M-002-022-MY3) (WWP, WTC), by the AI-MAT center of National Taiwan University (NTU-107L900802) and by the Natural Science and Engineering Research Council of Canada (JCB). Synthesis of the samples and some microscopic analyses were performed under the framework of RSF grant No 17-12-01519. The synchrotron X-ray diffraction beamtimes were awarded by National Synchrotron Radiation Research Center, Taiwan (Proposal Nos. S8\_1507\_17 and 2016B009-1). We thank Dr. M. W. Chu for fruitful discussion.
\end{acknowledgments}

\bibliographystyle{ieeetr}
\bibliography{./bib/references}

\end{document}